\documentclass[letterpaper,10pt]{article} 

\usepackage{osameet3} 

\usepackage{amsmath,amssymb}
\usepackage[colorlinks=true,bookmarks=false,citecolor=blue,urlcolor=blue]{hyperref} 
\usepackage{caption}
\usepackage{subcaption}
\usepackage[export]{adjustbox}
\newcommand*{\affmark}[1][*]{\textsuperscript{#1}}

\begin{document}

\title{Demonstration of a low latency bandwidth allocation mechanism for mission critical applications in virtual PONs with P4 programmable hardware}
\vspace{-4mm}
\author{Diego Rossi Mafioletti\affmark[1],\affmark[3], Frank Slyne\affmark[1], Robin Giller\affmark[2], Michael O'Hanlon\affmark[2], \\
David Coyle\affmark[2],
Brendan Ryan\affmark[2] and Marco Ruffini\affmark[1]}
\address{\affmark[1]CONNECT Centre, Trinity College Dublin, \affmark[2]Intel Corporation, Ireland, \affmark[3]Federal Institute of Esp{\'i}rito Santo}
\email{\{rossimad,fslyne,marco.ruffini\}@tcd.ie, \{robin.giller, michael.a.ohanlon, david.coyle,brendan.ryan\}@intel.com}

\copyrightyear{2021}

\vspace{-4mm}
\begin{abstract}
We provide a real-time demonstration of a low-latency PON DBA mechanism, optimised for virtual PONs. Our implementation mixes P4 programmable data plane and software-based virtual DBA to provide efficient fast-track allocation for low latency applications.
\end{abstract}
\vspace{-1mm}
\section{Overview}
The use of Passive Optical Networks (PONs) is a cost-effective approach for the delivery of ubiquitous broadband, since costs and capacity are shared across multiple end-points. This makes PONs ideal also for providing ubiquitous connectivity for next generation services and applications, delivering high performance at low cost.
The main issue is however that upstream scheduling works through a Dynamic Bandwidth Assignment (DBA) mechanism, which operates through a report/grant mechanism. This is much more efficient than static bandwidth allocation, however it introduces a latency of several hundred microseconds, which becomes an issue for application requiring end-to-end delay of 1 ms or less. This latency is further exacerbated when PONs are implemented as Virtual Network Functions, in modern implementations such as the SDN Enabled Broadband Access (SEBA) \cite{SEBA} and when using virtual DBA \cite{Ruffini2020}. While these bring the required flexibility to fully customise a PON for high-performance multi-service and multi-tenant operation, the physical separation between hardware transmission equipment and software scheduling algorithms can further increase upstream latency.

When the PON is used to provide low-latency fronthaul services to Cloud-RAN, DBA latency issues are solved by coordinating the 5G RAN and PON schedulers through a cooperative DBA algorithm \cite{coDBA}. In Cooperative DBA, the Distributed Unit (DU) in the RAN provides the OLT DBA scheduler with advance scheduling information of the packet arrival at the Remote Unit (RU) connected to the ONU. This enables the OLT to send the ONU a transmission grant that is available as soon as the wireless data packet is received by the RU/ONU, thus minimising upstream latency.

However, if the PON is used to carry data from an application that requires low latency (for example for remote control of robots and machinery), where packet arrival at the ONU cannot be determined in advance (or reasonably predicted), latency cannot be minimised by mechanisms such as the Cooperative-DBA.

In this paper we demonstrate a mechanism to significantly reduce PON latency in these second types of use cases,  when the application-level packet arrival at the ONU is unpredictable. The conceptual idea was initially introduced in \cite{P4_OFC}. The idea is to split the upstream DBA scheduling into two parts. The first part operates according to standard DBA procedures, where the bandwidth map is calculated after all required DBRus (i.e., the grant requests form the ONUs) are received for a given cycle. However, part of this bandwidth map is initially left unallocated. A second mechanism, which we call Fast Intercept, runs in the P4 hardware NIC, spoofing upstream low latency DBRu requests and storing them until the next bandwidth map (BWMAP) arrives from the downstream frame. It then modifies the BWMAP adding grants form the low latency DBRus on the unallocted portion of the BWMAP. The downstream frame with the BWMAP is then forwarded to all ONUs, following normal PON operations. 

The demonstration shows how this method, which is compatible with fully disaggregated PONs, can reduce latency significantly compared to standard DBA operations, as the low latency allocations are assigned through the BWMAP in the first available downstream XGEM frame. This mechanism is also much more efficient than fixed capacity allocation, as the dedicated capacity for low latency applications can be shared across all ONUs and be dynamically adjusted at every frame. 

\section{Innovation}
Data planes are generally rendered in hardware using  ASIC or FPGA, or in software running on a general purpose processor. The former can run at high line rates but are less flexible and require longer development time and effort. The latter runs at lower line rates, but are more flexible and require shorter development times. We have implemented a best of both worlds solution, where the portion of a scheduling algorithm that must operate at line rate can run on a programmable data plane, written in P4 language, while the portion that suffices for slower speed can run on the general purpose processor. Programmable data planes allow users to define their own data plane algorithm for network devices. This offers great flexibility for network customisation, be it for specialised, commercial appliances, e.g., in 5G or data centre networks, or for rapid prototyping in industrial and academic research. Programming protocol-independent packet processors (P4) is currently the most widespread abstraction, programming language, and concept for data plane programming. It is developed and standardised by an open community, and it is supported by various software and hardware platforms. P4 was developed to allow programmers to describe packet-processing functionality independently of the specifics of the underlying hardware. 

The interesting aspect we exploited in our demo is that P4 can also be used for the rudimentary processing of data. While it cannot perform the precise calculations possible with general-purpose central processing units and more complex programming languages, it can perform fixed-point arithmetic at line rate level speeds that are adequate for scheduling and resource allocation algorithms employed on passive optical networks. \textbf{The innovation of this work is the concept and implementation of a DBA scheduler that is split in two parts, one that can run on a P4-enabled network card and another, more complex, which runs on a general purpose processor}. The combination of the two into our Fast Intercept mechanism, provides the \textbf{critical trade-off between speed (i.e., low latency) and flexibility, as both data plane and vDBA scheduler have a high degree of flexibility}. 

Since the programmable P4 network device does not have the compute capacity for processing recursive and complex operations as a general-purpose processor, we use this device to only intercept latency-sensitive DBRu requests and carry out a simple, linear scheduling mechanism on top of an existing BWMAP. We bridge the gap between pure hardware DBA and virtualised DBA executed on the host.

\section{OFC Relevance}
Achieving low latency in the upstream direction is one of the biggest challenges in today's Passive Optical Networks. Together with capacity increase, it's one of the main requirements to enable the use of this flexible, low cost and potentially ubiquitous technology to support 5G and 6G services and applications. 
At the same time, network virtualisation and disaggregation is also a main component of modern networks, again providing flexibility, lower cost and better integration with heterogeneous network environments.

Our concept and demonstration brings together all these fundamental aspects, providing a flexible and fully programmable solution that achieves low latency on disaggregated Passive Optical Networks. For this reason, \textbf{we believe our demonstration is of interest to a wide audience, including attendees interested in optical access (especially PON) technology, network function virtualisation technology and software-defined control plane operations.} 
Our demonstration shows the Fast Intercept scheme that supports the virtualisation of the PON, while catering for low latency and low jitter applications that use PON as a bearer. Our scheme uses programmability both on data plane and on virtualised software platform that support customisation and multi-tenancy. 

The bulk of the scheduling calculations is executed as virtualised network functions on the host using Intel DPDK, however, key calculations are performed on the network interface card using a P4 language-capable programmable network device, shortening the path of the packet. 
\textbf{Attendees of the demo will witness innovative interaction between P4 and DPDK in a push-pull configuration, with packets bypassing the kernel network stack but are processed by user-land vDBA application.} 

\section{Demo content and Implementation}
\vspace{-4mm}
\begin{figure}[htp]
    \centering
    \begin{subfigure}[b]{0.37\textwidth}
        \centering
        \includegraphics[width=\textwidth,left]{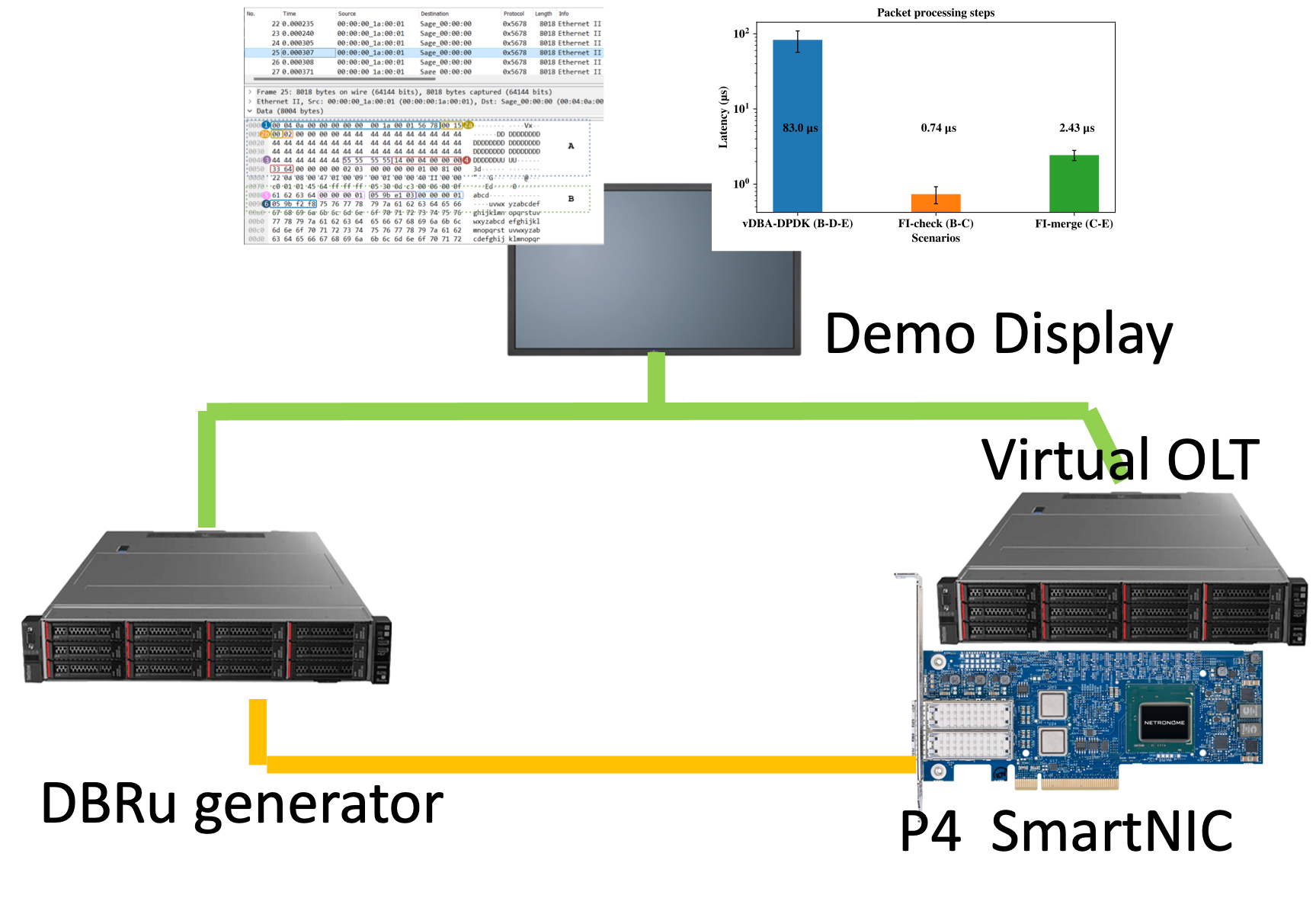}
         \vspace{-3mm}
       \caption{Physical demo Layout}
        \label{fig:layout}
    \end{subfigure}
    \begin{subfigure}[b]{0.57\textwidth}
        \centering
        \includegraphics[width=\textwidth,right]{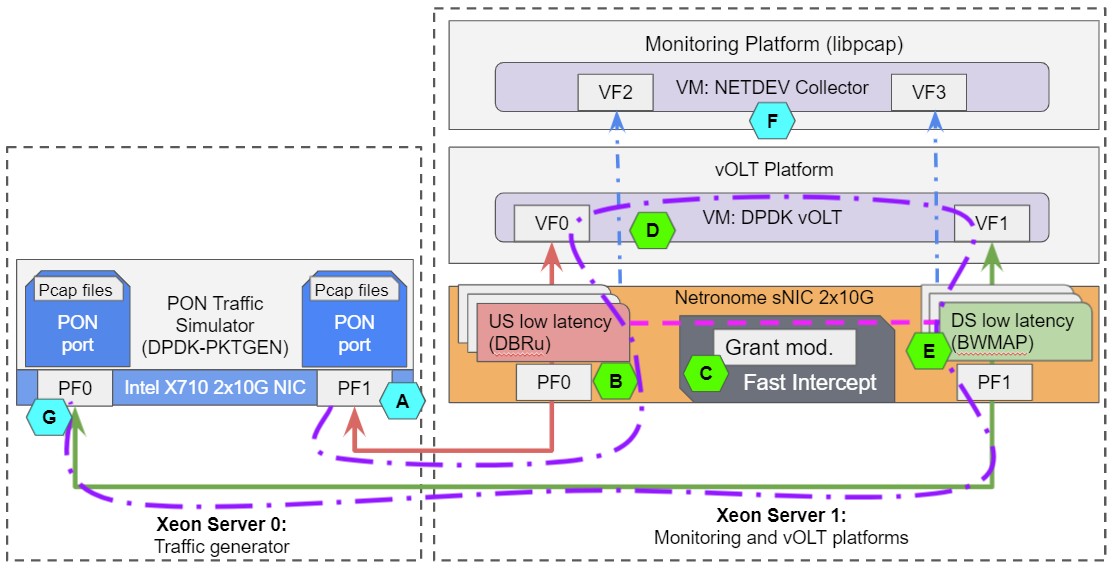}
        \caption{Logical view, showing DBA data paths}
        \label{fig:figure1}
        \end{subfigure}
        \vspace{-4mm}
    \caption{Logical and physical layout of the proposed demonstration}
    \vspace{-4mm}
\end{figure}

The demonstration will show the latency performance improvement of our Fast Intercept mechanism on a virtual PON, compared to standard DBA operations.
The physical setup is shown in Fig \ref{fig:layout}, with a server used to generate DBRu requests at 10G line rate (i.e., emulating multiple ONU operations) connected by a standard 10G NIC (Intel x710-DA2 Dual 10G Por) via fibre to the virtual OLT, which hosts the P4 Programmable SmartNIC (Netronome NFP-4000 Dual 10G port).
The corresponding logical setup is shown in Fig. \ref{fig:figure1}. Here two separate processes generate DBRu for normal traffic and low latency traffic, respectively, running on a DPDK-Pktgen-based traffic generator. These are intercepted by the P4 algorithm running in the SmartNIC. In the figure we can see that normal DBRu requests follow path (D), i.e., are relayed to the host processor for DBA calculation. The DPDK vDBA runs over a qemu hypervisor, accessing the SmartNIC virtual functions via SR-IOV. A single instance of the vDBA application runs on the system, based on the Intel DPDK software version 20.07, which runs on a 16-core 2.2 GHz Intel® Xeon® D-2100 processor-based server with 32 GB DDR4 DRAM.
Note that since this demo focuses on the scheduling algorithm, we don't implement a burst-mode receiver at the OLT. Thus the XGEM frames from the OLT are encapsulated over Ethernet before transmission. 

The BWMAPs are continuously generated by the DBA every frame and sent downstream. Here the SmartNIC runs an algorithm that intercepts the BWAMP in the downstream frame and includes allocations from the low latency DBRu requests that were last intercepted in the upstream frame.

\begin{figure}[htp]
\centering
\begin{minipage}{.65\textwidth}
\vspace{-2mm}
  \centering
  \includegraphics[width=\linewidth]{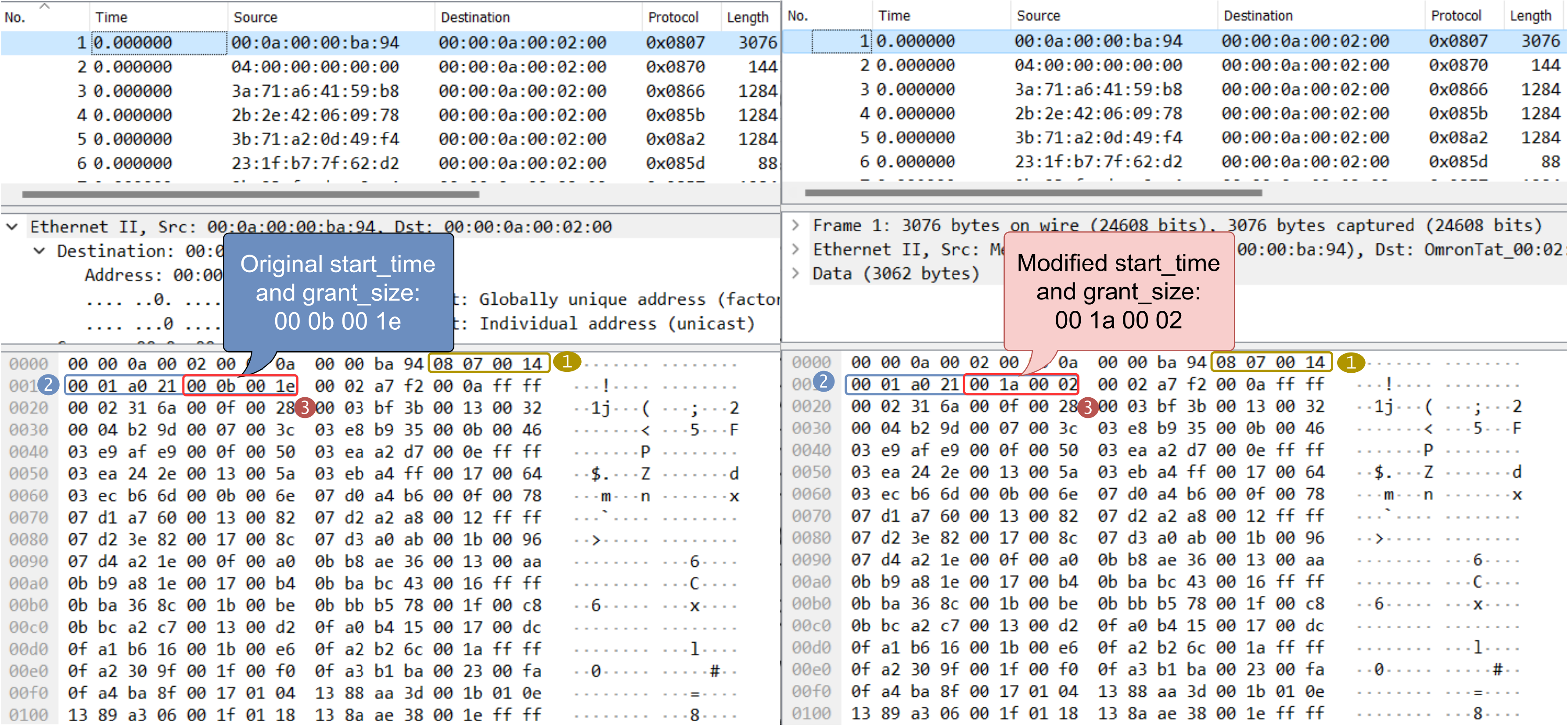}
  \vspace{-6mm}
  \captionof{figure}{BWMAP: original (left, blue text box) and merged/modified (right, red text box).}
  \label{fig:bwmap}
\end{minipage}%
\begin{minipage}{.35\textwidth}
  \centering
  \includegraphics[width=.95\linewidth]{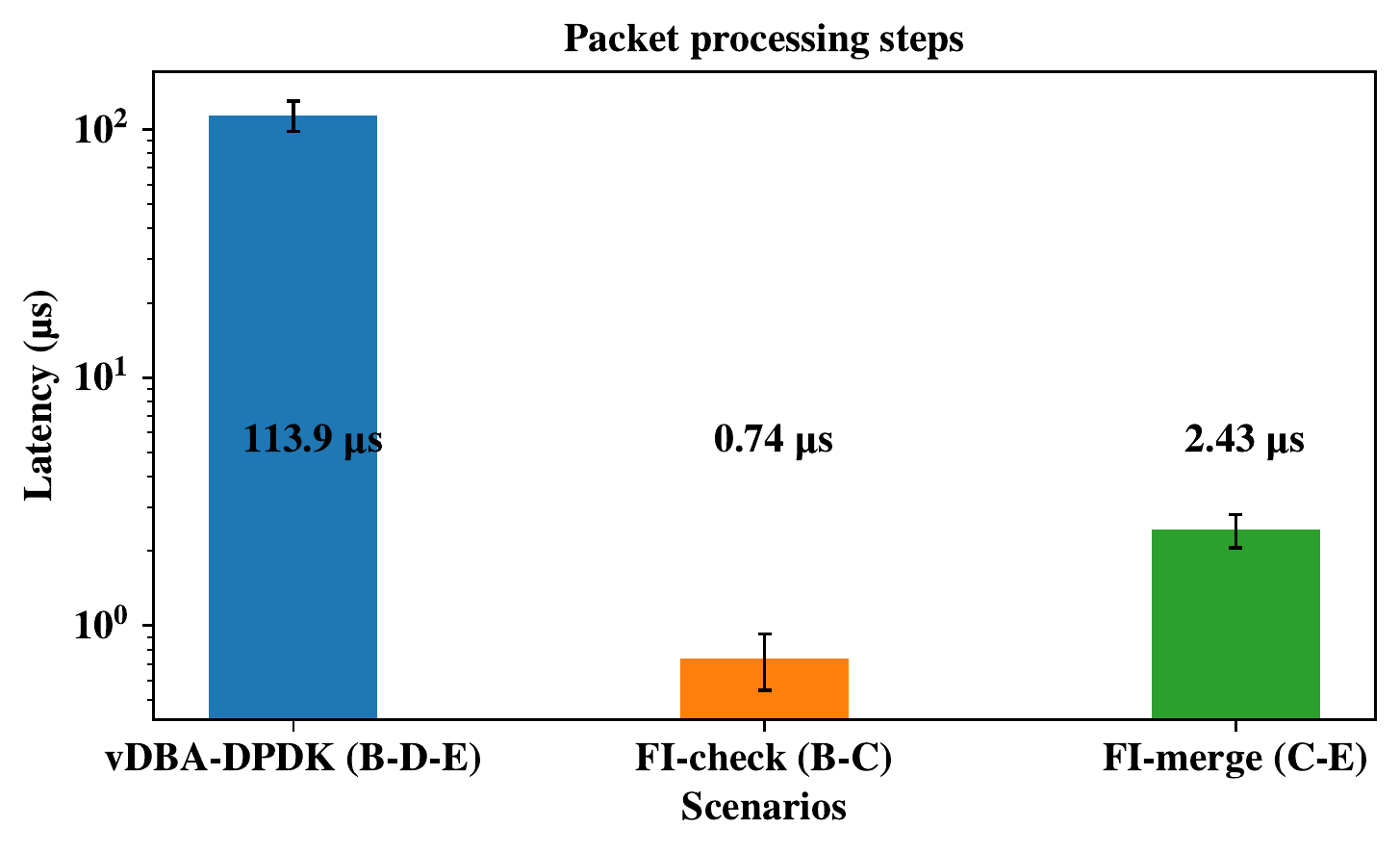}
  
  \captionof{figure}{Processing times}
  \label{fig:proctimes}
  \vspace{-8mm}
\end{minipage}
\end{figure}

During the demonstration, the following technical steps will be explained in detail. Considering the diagram in Fig. \ref{fig:figure1}, the P4 application in the SmartNIC (B) parses the DBRu data in the XGEM frame generated by the traffic generator (A). The DBRu data is routed according to the Allocation identifier (Alloc-ID) into host-based vDBA bandwidth allocation (D), for traditional network traffic, or Fast Interception (C) when a low-latency Transmission Container (T-CONT) is identified. The BWMAP is generated in both DPDK vOLT application and the Fast Intercept mechanism inside the SmartNIC. The Fast Intercept mechanism merges the original BWMAP with the new one (E), based on the Alloc-ID field, adjusting the grant allocations according to the traffic requirement. Concurrently, the Monitoring Platform (F) collects the packets for further inspection, storing them into PCAP files. During stages (B), (C), (D) and (E), the SmartNIC stores high-resolution hardware timestamps into the packets, to track the time to process and deliver the data through hardware and software levels. Each stage are calculated and displayed in the chart in Fig. \ref{fig:proctimes}, bringing us the time saved when using hardware for packet processing.

Figure \ref{fig:bwmap} shows the downstream frame, starting by the (1) HLEnd field and the specific (2) BWMAP generated by DPDK vOLT on the left side, following by the merged downstream frame, which was modified by the Fast Intercept mechanism, changing the BWMAP $start\_time$ and $grant\_size$ fields to a desirable value, ensuring the low-latency allocation to the specific T-CONT. 

\textbf{The demonstration will show this process running in real-time at line rate.} The attendees will first go through a slideset presentation, explaining the algorithm and showing a step-by-step guide to all operations. \textbf{At runtime, the demo will collect arrival information of the low latency and default packets, building a statistical representation of the latency distribution while the demo runs.} The attendees will thus be able to observe live application-level latency results, together with in-depth latency measurement of the algorithm's functions.



\vspace{-6mm}

\begin{thebibliography}{99} 
\bibitem {SEBA} S. Das. From CORD to SDN Enabled Broadband Access (SEBA). JOCN, Jan 2021.
\bibitem{Ruffini2020} M. Ruffini et al. Virtual DBA: virtualizing passive optical networks to enable multi-service operation in true multi-tenant environments. JOCN, Apr. 2020.
\bibitem{coDBA} T. Tashiro et al., A novel DBA scheme for TDM-PON based mobile fronthaul, Tu3F.3, OFC'14.
\bibitem{P4_OFC} D.R. Mafioletti, et al. A Novel low-latency DBA for Virtualised PON implemented through P4 In-Network Processing. OFC'21. 


\end{thebibliography}
\end{document}